\documentclass[letterpaper]{article} 
\usepackage{aaai25}  
\usepackage{times}  
\usepackage{helvet}  
\usepackage{courier}  
\usepackage[hyphens]{url}  
\usepackage{graphicx} 
\usepackage{natbib}  
\usepackage{caption} 
\usepackage{algorithm}
\usepackage{algorithmic}

\usepackage{newfloat}
\usepackage{listings}
\DeclareCaptionStyle{ruled}{labelfont=normalfont,labelsep=colon,strut=off} 
\floatstyle{ruled}
\newfloat{listing}{tb}{lst}{}
\floatname{listing}{Listing}
\title{Retrieval Augmented Generation-Based \\Incident Resolution Recommendation System for IT Support}

\author {
    Paulina Toro Isaza,
    Michael Nidd,
    Noah Zheutlin,
    Jae-wook Ahn,
    Chidansh Amitkumar Bhatt,\\
    Yu Deng,
    Ruchi Mahindru,
    Martin Franz,
    Hans Florian,
    Salim Roukos
}
\affiliations {
    IBM Research\\
    ptoroisaza@ibm.com, 
    mni@zurich.ibm.com, 
    noah.zheutlin@ibm.com,
    jaewook.ahn@us.ibm.com,
    chidansh.amitkumar.bhatt@ibm.com,
    dengy@us.ibm.com,
    rmahindr@us.ibm.com,\\
    franzm@us.ibm.com,
    raduf@us.ibm.com,
    roukos@us.ibm.com
}

\usepackage{bibentry}

\begin{document}

\maketitle

\begin{abstract}
Clients wishing to implement generative AI in the domain of IT Support and AIOps face two critical issues: domain coverage and model size constraints due to model choice limitations. Clients might choose to not use larger proprietary models such as GPT-4 due to cost and privacy concerns and so are limited to smaller models with potentially less domain coverage that do not generalize to the client's domain. Retrieval augmented generation is a common solution that addresses both of these issues: a retrieval system first retrieves the necessary domain knowledge which a smaller generative model leverages as context for generation. We present a system developed for a client in the IT Support domain for support case solution recommendation that combines retrieval augmented generation (RAG) for answer generation with an encoder-only model for classification and a generative large language model for query generation. We cover architecture details, data collection and annotation, development journey and preliminary validations, expected final deployment process and evaluation plans, and finally lessons learned. 
\end{abstract}

%

\section{Introduction}
The recent boost in performance and popularization of generative models has resulted in clients across various domains requesting generative AI powered question-answering and recommendation systems. However, there are two critical issues facing many clients wishing to implement generative AI: domain coverage and model size constraints due to model choice limitations. Much of the focus for generative models has been on the general domain and only some specific tasks such as coding. Models that work on these domains might not necessarily work for a client's targeted domain. Additionally, clients might choose not to leverage larger proprietary models such as GPT-4 because of cost and privacy concerns so clients are limited to smaller models with less domain coverage and likely lower out-of-the-box performance. 

We have faced both of these issues when building a solution recommendation system for resolving IT support cases. Models and tasks within the domain of IT support and  Artificial Intelligence for IT Operations (AIOps) are under-researched. No IT support specific fine-tuned generative AI model exists and there are limited publicly available datasets on IT support tasks like question-answering (QA) or retrieval over a corpus of IT support documents. Thus, it is non-trivial to evaluate out-of-the-box AI models on IT support tasks as well as train and evaluate custom AI solutions in this domain. 

In specific, the IT support
use case involved IT product support tickets that are opened by a customer and answered by a support agent after a series of interactions. 
It required a system that would respond only to cases that did not need any additional information or clarification from the customer. That is, the support case could be resolved based only on the initial information present in the case subject and description when the ticket was first opened. Additionally, the use case required that solutions be grounded in an official support document that would be presented to the customer as a link along with the summarized solution. 

Given these use case requirements, retrieval augmented generation (RAG) was a natural fit \cite{10.5555/3495724.3496517}. It is a common solution that addresses the two main issues of inadequate generalizability to more niche domains and model size limits. It does so by using a retrieval system to first retrieves the necessary domain knowledge which a smaller generative model then leverages as context for answer generation. 

We present the resulting system for IT support case solution recommendation that combines an encoder-only model for classification, two dense embedding models for retrieval, and generative large language models for query and answer generation. The solution brings several novel contributions to the field of AIOps:
\begin{itemize}
  \item The first reported evaluation of a RAG system for IT support incident resolution recommendation.
  \item The first reported use of a classifier for determining single vs multi-turn IT support cases.
  \item Evidence of substantial retrieval improvement using re-ranking based on a new model, IBM Slate 125m \cite{Slate-125}
  \item A comparison of answer generation performance across diverse model sizes that shows smaller models can match or even beat the performance of very large models in the RAG incident remediation use case.
\end{itemize}

\section{Architecture}
\begin{figure}
\centering
\includegraphics{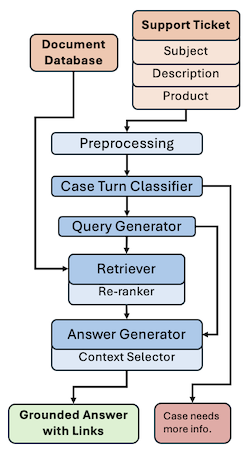}
\caption{System architecture}
\label{fig:arch}
\end{figure}

Given a support case subject, description, and product name, our system generates recommended solutions based on corpora of support documents. Our system consists of four major components as illustrated in Figure \ref{fig:arch}: an encoder-only transformer \textit{classifier}, a \textit{query generation} system, a \textit{retriever} system, and an \textit{answer generator} system. 

\textit{\textbf{Preprocessing:}} Support cases are ingested with unstructured text data fields of case subject, case description, product name, and product version number. The escape and non-ASCII characters are removed from the case subject and description, and the two fields are concatenated. We preprocess the product name by matching it to a dictionary of known product acronyms or alternative names to append to the query used in retrieval. 

\textbf{\textit{Case Turn Classifier:}} The cleaned and concatenated case subject and description are fed into the classifier which determines if the support case is a single turn. ingle-turn cases are defined as those that can be resolved using only the information present in the case subject and description, without requiring any additional information or clarification from the customer. The classifier is an encoder-only transformer model IBM Slate 125m \cite{Slate-125} that was fine-tuned on almost 14,000 examples labeled by subject matter experts. 
If the case is predicted to be single turn, the case continues to the next step in the pipeline. 

\textbf{\textit{Question Generator:}} The question generator summarizes the often vague case subjects and verbose, convoluted descriptions into concise text queries suitable for the retrieval system. The system prompts a large language model, Mixtral 8x7B Instruct \cite{jiang2024mixtralexperts}, to generate a concise question based on the provided case subject and description. Additional post-processing keeps only the first generated sentence in case the generative model does not follow instructions and generates additional sentences beyond the first question.

\textit{\textbf{Retriever:}} Our documentation is retrieved from multiple data collections in a Milvus vector database, all indexed with the standard Slate-30M embedding. If the search stage requires top-3 documents, we search for 3
from each of these indexes, and then merge-sort based on the score before retaining the top-$k$ from the combined set.

Having retrieved the top-3 documents, as ranked by a
general-purpose embedding, we re-rank them using a Bi-Encoder model 
that has been fine-tuned using application-specific training data.
Re-ranker scores are computed as cosine similarities of a combination of the original case and the derived question, compared with the same passages (with recalculated vectorizations) that were matched in the first pass.

\textit{\textbf{Answer Generator: }}For each of the top three documents retrieved, the answer generator produces an answer to the previously generated query by leveraging the top three re-ranked passages from the document. First, we split the retrieved document content into 2500-token chunks with 250-token overlaps and use cosine similarity to pick the three most relevant chunk contexts. The answer generator prompts a large language model, Mixtral, to answer the query using the three contexts. Finally, the system returns the three retrieved links with their corresponding generated answers. The results can then be displayed to the support agent in a graphic interface. 

\section{Data} 
\label{sec:data}
We collected almost 19,000 real support cases across nine software products: six to serve as training and three to serve as validation. Each case included a case subject and description originally drafted by a customer when opening the case. Additionally, the three indices for documentation leveraged for retrieval had a total corpus of over 5 million documents. 

For each product, we asked five support agents who were singled out as product subject matter experts to carry out the following tasks for each of the nine products: 1) annotate single-turn vs. multi-turn label, 2) validate silver ground truth query based on case subject and description and provide updated query as necessary, 3) provide link to relevant document in support corpus, and finally 4) copy and paste solution to query as found in the relevant support document.

Tasks \#2 through \#4 were carried out only for cases that had been labeled as single-turn. After cleaning and removing missing annotations, we created a dataset of almost 19,000 support cases, with almost 3,000 cases for each training product and over 400 cases for each evaluation product. 

\section{Development and Validation}
\label{sec:dev_val}
\subsection{Case Turn Classifier}

As our solution is meant to supply answers before involving a support agent, we needed to develop a method for classifying incoming cases as single-turn vs. multi-turn. The classifier model is a binary encoder-only IBM Slate 125m model fine-tuned on the single-turn/multi-turn labels of almost 7,000 unique cases across six software products. The final training data is around 14,000 cases as it includes two copies of a given case: one with tokens from both the case subject and description fields, and another with tokens only from the case subject..

\begin{table}[h]
\fontsize{9pt}{9pt}
\centering
\setlength{\tabcolsep}{1mm}
\begin{tabular}{l|rrrrrr}
 &  &  & \textbf{Positive} & &  &  \\
 & \textbf{Train} & \textbf{Eval} & \textbf{Class \%} & \textbf{F1} & \textbf{P} & \textbf{R}    \\
\hline
In-Domain & 13964  & 3512 & 25\% & 0.46 & 0.31 & 0.89 \\
Out-of-Domain &  - & 1375 & 53\% & 0.65 & 0.54 & 0.80
\end{tabular}
\caption{Single-Turn classifier model performance on six in-domain training products and three out-of-domain evaluation products}
\label{tab:val_class_train_products}
\end{table}

For our application, we prioritized correctly predicting single-turn cases (positive class) versus multi-turn cases (negative) emphasizing recall over precision. Table \ref{tab:val_class_train_products} presents the final fine-tuned model performance on the held-out evaluation set of the six products when using a classification threshold of 0.1. We found that performance varies widely depending on the product, ranging from F1 of 0.27 to 0.62 and recall from 0.75 to 0.98. The lower performance can be explained in part by the varying class imbalance across products (positive class proportion from 11\% to 44\%) as well as the products' differing inter-annotator agreement (See Section 6). Despite this, the model still substantially outperforms random guessing of the classes. Hyper-parameters including batch size, learning rate, and dropout were determined based on a small grid search. 

We then evaluated the fine-tuned classifier on about 1,400 cases from three additional products that were not in the training set to validate if the model generalized to other products. The resulting F1 of 0.65, precision of 0.54, and recall of 0.80 for the three products suggests that classifier model generalizes well to products not seen during training even with substantially different class balance.

\subsection{Query Generator}

\begin{table}[h]
\fontsize{9pt}{9pt}
\centering
\resizebox{1.00\columnwidth}{!}{
\begin{tabular}{l|rr}
\textbf{Model} & \textbf{BertScore F1} & \textbf{ROUGE-L F1} \\
\hline
Falcon-40B* & 0.91 & 0.40 \\
Mistral-Large-2 & 0.91 & 0.38 \\
Mixtral-8x7B-Instruct & 0.91 & 0.36 \\
Granite-13B-Chat-v2 & 0.89 & 0.28 \\
\end{tabular}
}
\caption{Comparison of BertScore F1 and ROUGE-L F1 for different models on query generation task. BertScore is based on roberta-large embeddings. \\
*Falcon-40B scores are not comparable to the other models because the prompt used was different, and it was used to create the initial questions that were validated or edited by SMEs to create the ground truth.}
\label{tab:val_query}
\end{table}

In order to create a concise query that could be used by the retriever, we generated a single sentence question based on the case subject and description. Our experiments over various open-source generative models (Table \ref{tab:val_query}) and model availability in the client's services led us to choose Mixtral-8x7b-Instruct as the model for query generation which reliably reproduced the ground truth queries despite being a relatively small model with no domain knowledge. Note that the results are skewed for Falcon-40B \cite{falcon40b} as Falcon-40B generated the first pass of silver ground truth queries that were then edited by subject matter experts. 

\subsection{Retriever} \label{subsec:retrieval}
Support experts supplied us with a collection of cases with one ground truth link each that a ``correct'' solution should reference; our evaluation is based on whether this link is contained in the top $n$ links returned (for various values of $n$).  Implementing this evaluation presented several challenges:
\begin{itemize}
\item \textit{URL Duplication:} A single page of documentation often has several different URLs to identify it.
\item \textit{Subtle Content Variations:} Documentation for the same topic in different versions of a product may have subtly different titles, like ``How to update a list'' and ``Lists: Updating''.
\item \textit{Identical Content Across Different Documents:} Documentation for the same topic in different versions may be identical in which case results from different versions are still valid.
\end{itemize}

Mitigating the first of these challenges, many of our documentation pages include a ``canonical link'' in their metadata.
In many cases, this allows us to identify identical links.
The two issues with documentation evolution between versions are addressed with Rouge-1 scores, using a threshold of 0.90 as sufficiently similar to count as identical.

For retrieval, a dedicated team is already responsible for maintaining indexed collections of the software product documentation.
This saves our project from gathering, maintaining, and indexing all of these documents, and its base Slate-30M embedding returns a good first set of results.

\begin{figure}
    \centering
    \includegraphics[width=0.95\linewidth]{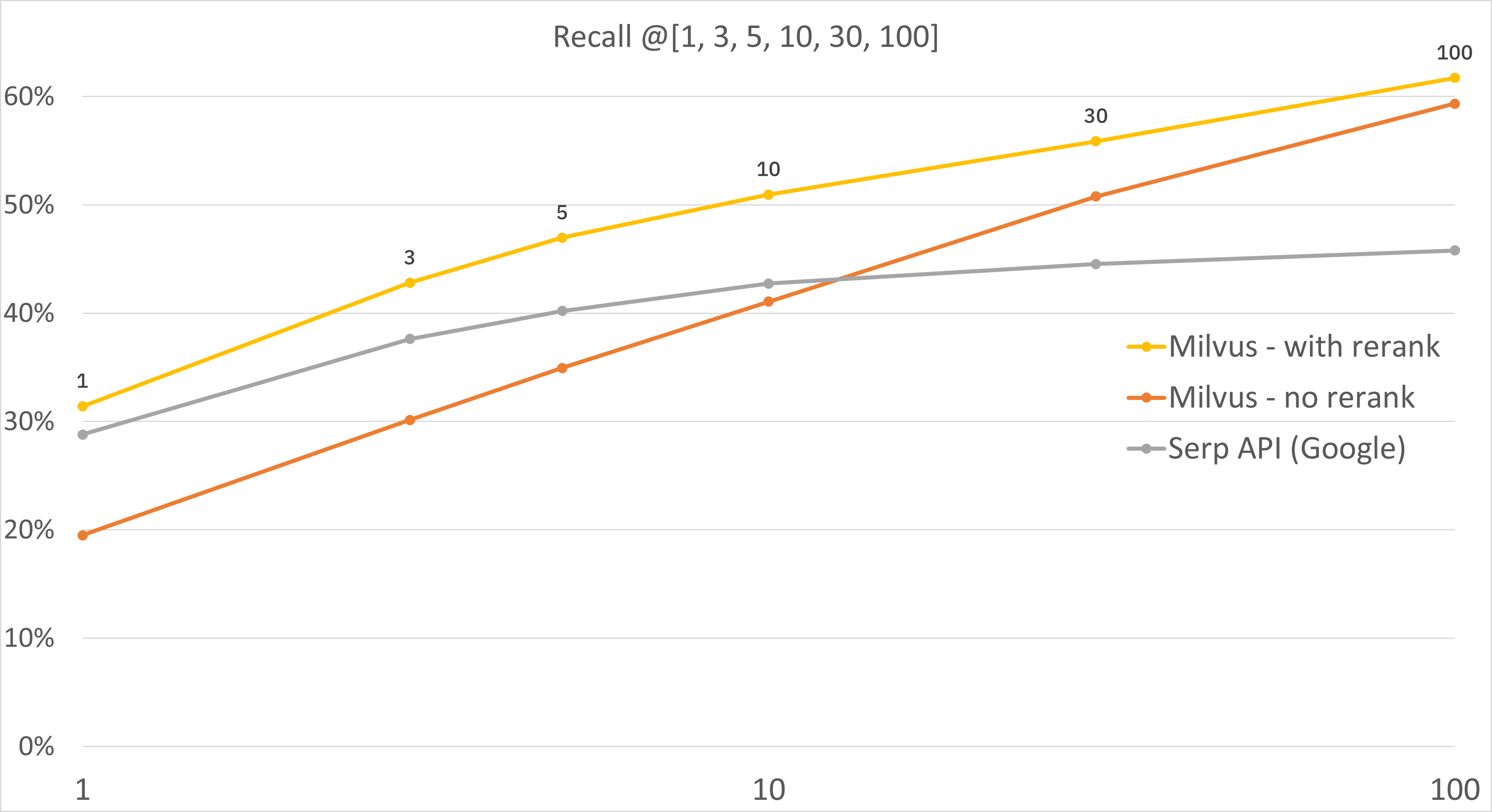}
    \caption{Retriever recall for top $x$ (log scale), comparing with (yellow) and without (orange) re-ranking vs. direct Google search via SerpApi (grey) for 1729 customer issues over six products}
    \label{fig:recall-rates}
\end{figure}

Re-ranking this first set allows us to use fine-tuning to improve performance without maintaining a parallel set of indexes.
For this final task-specific fine-tuning stage, we used training data based on 1,430 questions with up to three matching passages per question extracted from documents identified in user interactions, together with negative examples found using BM25 search.  The resulting IBM Slate-125M model was then distilled into the deployed IBM Slate-30M model. To evaluate its effectiveness, we present recall before and after re-ranking with Google Search as a baseline in Figure~\ref{fig:recall-rates}.

\begin{table}[ht]
\fontsize{9pt}{9pt}
\centering
\setlength{\tabcolsep}{1mm}
\begin{tabular}{l|ccc}
& \textbf{0 - Completely} & \textbf{1 - somewhat}     & \textbf{2 - Solution} \\
\textbf{Rating} & \textbf{irrelevant} & \textbf{relevant/helpful}     & \textbf{in link} \\
\hline
Product A \\
SME & 58\% (11) & 0\% (0) & 42\% (8) \\
Tool & 26\% (5) & 5\% (1) & 68\% (13) \\
\hline
Product B \\
SME & 20\% (12) & 48\% (28) & 28\% (17) \\
Tool & 48\% (29) & 37\% (22) & 12\% (7) \\
\end{tabular}
\caption{AB testing human evaluation of retrieved links}
\label{tab:poc_rating}
\end{table}

We also conducted an AB test in which support agents of two products were provided with a link retrieved by the tool and a link provided by a subject matter expert. The source of the link was randomized as Source A or Source B so that, for example, Source A could be either our tool or an SME for any given case. The support agents were asked to rate each link as shown in Table \ref{tab:poc_rating} and to pick the better of the two solutions. 

The results for Product A show that support agents gave higher ratings to and preferred the links suggested by the tool over those from a SME. The results for Product B however show higher scores for links provided by the SME but about half of the cases were still rated as having a somewhat helpful or complete link provided by our tool. Additionally, when directly asked to compare the recommendations, support agents reported that the tool was more helpful or just as helpful as the SME link 69\% of the time for Product A and 35\% of the time for Product B. 

\subsection{Answer Generator}
The final step of our solution takes in the generated query and top three most relevant retrieved passages as context to prompt the answer generator. In particular, the prompt asks the model to use the information within the provided contexts to generate an answer. Additionally, if the context is insufficient, then the model is instructed to state that an accurate answer cannot be provided.

To evaluate the answers, we used the subject matter expert's annotated ground truth answers and ground truth documents verified to contain the answer to the question. We compared the answers generated by the answer extractor using the ground truth document to the ground truth answer using BertScore \cite{zhang2020bertscoreevaluatingtextgeneration} and ROUGE-L F1. We evaluated different models and prompts to find the optimal combination and present the results of the models assessed in Table \ref{tab:answer_generator}. While BertScore (roberta-large) F1 is rather low (in practice it ranges between 0.85-0.95), ROUGE-L F1, traditionally a rather strict metric, shows promising results for Mixtral-8x7b-Instruct with a score of 0.41. Mixtral-8x7b-Instruct's outperforms of GPT-4o, included as a baseline for larger models, in all three metrics, despite having substantially less parameters. Likewise, Granite-13B-Chat-v2 is not far behind GPT-4o despite its merely 13 billion parameters compared to GPT-4o's rumored hundreds of billions or even trillions of parameters. This suggests that the RAG approach of smaller models leveraging retrieved context is a viable solution for IT incident resolution recommendation systems. 

\begin{table*}[ht!]
\centering
\begin{tabular}{l|ccc}
\fontsize{9pt}{9pt}
\textbf{Model} & \textbf{BertScore (roberta-large) F1} & \textbf{BertScore (deberta-xlarge-mnli) F1} & \textbf{ROUGE-L F1} \\
\hline
GPT-4o (2024-08-06) & 0.86 & 0.62 & 0.34 \\
Mistral-Large-2 & 0.86 & 0.62 & 0.37 \\
Mixtral-8x7B-Instruct & 0.87 & 0.64 & 0.41 \\
Granite-13B-Chat-v2 & 0.86 & 0.58 & 0.32 \\
\end{tabular}
\caption{Comparison of BertScore F1 and ROUGE-L F1 for different models performing the answer generation task. BertScore based on roberta-large embeddings}
\label{tab:answer_generator}
\end{table*}

\textbf{\textit{Knowledge Infusion for Answer Generation}}

Directly applying general foundational models to AIOps for answer generation tasks often does not yield optimal results. The knowledge infusion approach involves adapting these pre-trained models to specific tasks through additional training on task-specific data. 

To further boost the results of our action generation task for AIOps domain, we employ the knowledge infusion methodology described in \cite{sudalairaj2024lablargescalealignmentchatbots}. 
First, we manually created a seed dataset with six tuples, each containing context and four related question-answer pairs. Then, we randomly selected fifteen documents from the corpus to guide synthetic data generation, using the seed dataset to replicate similar artifacts for each document. Using this seed dataset, we created 14,000 synthetic samples with the Mixtral-8x7B-Instruct model as the teacher. IBM’s Granite 7B IL-Internal-Granite-7B-Base, a much smaller model, was fine-tuned with IT domain data to cater to the specific task of answer generation for the IT Support use case resulting in the InstructLab-IT model with domain knowledge infusion.

To evaluate the quality improvement, we conducted a user study with 6 technical experts and forty test question-answer pairs per model. For each question, we retrieved context from a 1200-document corpus of six software products and used it to prompt each model separately for an answer. Our user study used a 0 to 1 rubric to evaluate answer correctness with clear descriptions for each level: 
\begin{itemize}
\item 0 = Incorrect: irrelevant or fails to answer the question.
\item 0.25 = Mostly Incorrect: Some details are correct, but key details are missing, fabricated, or mostly irrelevant.
\item 0.5 = Partially Correct: Most details are correct, but some key details are missing, fabricated, or include a lot of irrelevant information.
\item 0.75 = Mostly Correct: Most details are accurate, with only minor gaps or irrelevant information.
\item 1 = Correct: Includes only relevant details. 
\end{itemize}

In Table \ref{tab:val_rubric}, we present the final score for each model as the average of human annotators' scores across 40 question-answer pairs of which InstructLab-IT emerged as the best model. While Llama-3.1-8b-Instruct performed slightly better than GPT-4o, the improvement in results with InstructLab-IT was very noticeable over both models. This is especially significant considering the model sizes: GPT-4o (over 1 trillion parameters and 1.5 TB), Llama-3.1-8b-Instruct (8 billion parameters and 16 GB), and InstructLab-IT (7 billion parameters and 28 GB). These results signal that a smaller, domain-specific model tuned for a specific set of use cases may better meet client requirements.

\begin{table}[h]
\centering
\begin{tabular}{l|r}
\fontsize{9pt}{9pt}
\textbf{Model} & \textbf{Score} \\
\hline
GPT-4o & 0.68  \\
Llama-3.1-8b-Instruct & 0.70 \\
InstructLab-IT & 0.76 \\
\end{tabular}
\caption{Comparison of analytic rubric scores for different models on answer generation task.}
\label{tab:val_rubric}
\end{table}

\section{Deployment}
\label{sec:deploy}
\begin{figure*}
  \includegraphics[width=\textwidth]{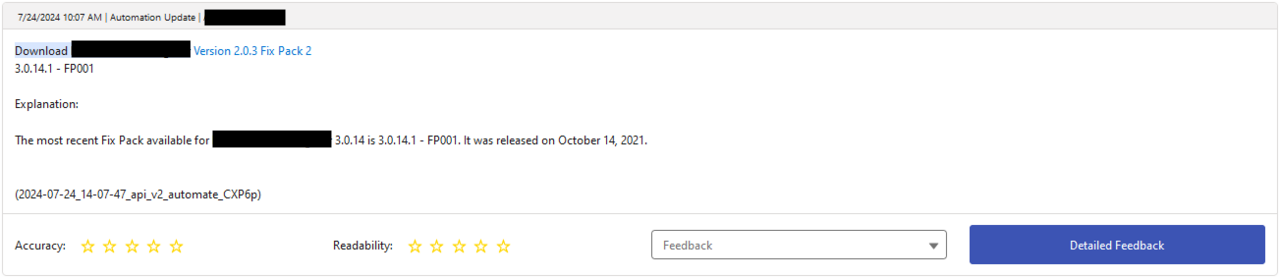}
  \caption{Working mockup of online deployment UI of single system result with feedback items.}
  \label{fig:feedback}
\end{figure*}

The tool is currently integrated into the ticketing system but silently deployed (not visible to agents) for testing purposes. We are currently working on refinements and integration and plan to deploy the system before the end of the year. We will incorporate feedback buttons for the tool once it is deployed online and visible to agents. In the customer support portal user interface, for a given case, support agents will see a suggested solution and link. This will include five-star ratings for accuracy and readability as well as a drop-down menu for feedback including the options: ``useful", ``somewhat useful", ``no useful suggestion", and ``need more client info". See Figure \ref{fig:feedback} for a working mock-up of the user interface. 

\section{Lessons Learned}
\label{sec:lessons}
\subsection{Use Case Formation}
\label{lessons:use_case}
Defining the proper use case is probably the most critical step in developing a proper RAG recommendation application. Because of the expense of data collection, annotation, and development, any confusion or change in the exact use case and capabilities of the model can result in substantial delays and costs. 

For example, the first dataset that we considered for this use case was synthetic data for which subject matter experts crafted questions based on support document titles, and then provided corresponding answers.  When we compared this data to actual customer cases, we found the genuine questions to be more verbose and to contain more off-topic ``noise.'' Thus we decided to use the more challenging actual support ticket data for training and validation, as it appeared better suited to our final deployment than the cleaner synthetic data.

We recommend spending time early on to understand how stakeholders will interact with the system, knowing that changes and evolution in the actual workflow may cause a decrease in system performance. 

\subsection{Inter-Annotator Agreement}
\label{lessons:agreement}

\begin{table}[h]
\centering
\begin{tabular}{l|ccc}
\textbf{Product} & \textbf{Classifier}     & \textbf{Question}          & \textbf{Link}       \\
\hline
Product 1     & 0.80 (20)      & 0.75 (16)         & 0.50 (16)  \\
Product 2     & 0.50 (20)      & 0.50 (10)         & 0.70 (10)  \\
Product 3  & 0.15 (20)      & 1.00 (3)          & 1.00 (3)   \\
Product 4      & 0.65 (20)      & 0.85 (13)         & 0.69 (13)  \\
Product 5     & 0.65 (20)      & 0.54 (13)         & 0.69 (13)  \\
Product 6    & 0.25 (20)      & 1.00 (4)          & 1.00 (4)   \\
\hline
Total   & 0.50 (120)      & 0.72 (59)         & 0.67 (59)    
\end{tabular}
\caption{Inter-Annotator Agreement: Proportion of labels that 3 annotators agreed on. Total N in parenthesis. For question and link labels, proportions only calculated based on cases for which all 3 annotators labeled as single turn and evaluated quality of corresponding question and link.}
\label{tab:agree}
\end{table}

Three SMEs labeled a subset of twenty cases to determine inter-annotator agreement. The results in Table \ref{tab:agree} show that labeling cases as single vs. multi-turn is not a trivial task and for most products, SMEs disagreed widely. Of the cases in which all three annotators agreed to be single-turn, agreement on the question and link quality was better but still raises questions about the validity of the training and evaluation data. In particular, the low agreement of the provided links can be explained by the fact that more than one link can potentially solve the same question and so neither annotator is necessarily wrong. This suggests that for ground truth data, we should consider a list of correct links instead of a single ground truth link for each question. The low agreement of single-turn vs. multi-turn labels also potentially explains the lower performance of the classifier model if the model is attempting to learn from potentially conflicting information.

\subsection{RAG Bottlenecks}
\label{lessons:rag}

The major bottleneck in RAG systems is the retrieval component. As shown in Table \ref{tab:answer_generator}, when given the correct context, LLMs can typically generate responses that match the ground truth answers. However, we cannot expect to generate the correct answer if given the wrong contexts which happens for around 60\% of the cases (Figure \ref{fig:recall-rates}). For comparison, Google search limited to the corresponding domains indexed by the Milvus database performed worse at 30\% R@3 compared to our method at 43\% R@3 (See Figure \ref{fig:recall-rates}). This implies, as other researchers have suggested, that the retrieval component in RAG is not a solved problem by any means. \cite{2024irrag, 2024redefiningretrieval}

\section{Related Work}
\subsection{LLM-Based AIOps}
As software systems become more complex, Artificial Intelligence for IT Operations (AIOps) methods are widely used to manage software system failures and ensure the high availability and reliability of large-scale distributed software systems \cite{Zhang2024ASO}. Machine learning and natural language processing methods such as LLMs have been used in AIOps for incident triage, data pre-processing, failure perception, root cause analysis, and auto remediation \cite{Zhang2024ASO}. Historically and currently, many of these tasks including both incident triage and auto remediation have been treated as classification problems: for example, \citet{AhmedKB2023} treats incident resolution as a classification task matching incident tickets to a relatively small number of possible resolutions using the BERT model and embeddings. With the rise of better performing generative AI models, researchers have moved towards using these models to generate solutions in the auto remediation task using prompting strategies \cite{AhmedLLMRcaM2023, liu2024opsevalcomprehensiveoperationsbenchmark} or creating a model fine-tuned for a variety of IT tasks such as question-answering \cite{Guo2023OWLAL}. 

Our use case can be considered an example of \citet{Zhang2024ASO}'s "Assisted Questioning", an auto remediation task that involves utilizing LLMs to aid operations personnel in answering system-related queries. 
As far as we are aware, no current work exists that leverages a RAG-based approach to solve this task, although one does exist for a similar IT task of root cause analysis \cite{ChenRCAPilot2024}. The RAG-based approach was taken in lieu of fine-tuning such as in \citet{Guo2023OWLAL}'s OWL model because of issues in real-world deployment due to its resource-intensive nature which requires significant computational resources and the interpretation of model decisions. Likewise, we discounted using a simple prompting approach without retrieval because of client limitations in model choice that prevented us from using larger models.


\subsection{Retrieval and Retrieval Augmented Generation}
Methods for finding the relevant documents or passages to answer a user query are typically divided into sparse~\cite{robertson2009} and dense retrieval systems~\cite{Zhao2024}. Our retrieval starts with a Milvus~\cite{milvus-definition} vector database that has indexed the 
software support documentation with a general purpose \textit{dense} embedding that serves multiple services.
In our solution, we then make use of a popular optimization by re-ranking the first-pass result to obtain a more appropriate ranking for our particular application \cite{nogueira-cho-2020, han-et-al-20}, giving us the results of a special-purpose index while still retaining the benefits of a central indexing service. Other popular improvement methods include combining sparse and dense embeddings into a hybrid system \cite{luan-etal-2021}.

Retrieval augmented generation was developed to address cases in which large language models have not learned and stored domain knowledge through pre-training. Originally implemented by combining dense retrieval and a fine-tuned BART model, the method generalizes well to larger generative models \cite{fan2024surveyragmeetingllms}. The quality of the answer generation can be improved through prompt engineering, refining how the specific generative model is prompted with the context, question, and other instructions. \cite{Liu2023}. However, it has been noted that the retrieval component in RAG has been understudied in comparison to the generation component despite its substantial impact on the final performance of such hybrid systems. \cite{2024irrag, 2024redefiningretrieval} 

\section{Conclusion}
We were able to deliver a first working version of an IT product solution recommendation system employing RAG. To our knowledge, this is the first published architecture and performance metrics of such a RAG system in this domain. Our system also differentiates itself with a few innovations including a component for classifying support cases as single-turn and a component for distilling verbose case descriptions into a query suitable for retrieval. We demonstrate that smaller models leveraging retrieved domain context can match or out-perform substantially larger models both with and without context \cite{AhmedLLMRcaM2023, liu2024opsevalcomprehensiveoperationsbenchmark} particularly with knowledge infusion through fine-tuning. However, there are still many challenges in implementing RAG for IT support incident resolution including improving retrieval performance. We are collecting feedback from support specialists using our current deployment, and intend to incorporate their advice into future improvements. 

\clearpage

\bibliography{aaai25}

\end{document}